\documentclass[aps,pre,showpacs,reprint,groupedaddress,amsmath,amssymb,sectsty,cite,url]{revtex4-1}

\usepackage{epsfig}
\usepackage{color}

\definecolor{graphbach1}{rgb}{0.5,0,0}
\definecolor{graphbach2}{rgb}{0,0.5,0}
\definecolor{graphbach3}{rgb}{0,0,0.5}
\usepackage[section]{placeins}

\newcommand{\be}{\begin{equation}}
\newcommand{\ee}{\end{equation}}

\newcommand{\ba}{{\bf a}}

\newcommand{\br}{{\bf r}}
\newcommand{\bu}{{\bf u}}

\newcommand{\bx}{{\bf x}}

\newcommand{\bF}{{\bf f}}
\newcommand{\bA}{{\bf A}}

\newcommand{\bU}{{\bf U}}

\newcommand{\tbx}{{\wt{\bf x}}}
\newcommand{\bW}{{\bf W}}
\newcommand{\rmd}{{\rm d}}
\newcommand{\dt}{{\frac{\rmd}{\rmd  t}}}

\newcommand{\tbr}{\tilde{\bf r}}

\newcommand{\wt}{\widetilde}

\newcommand{\brho}{{\mbox{\boldmath $\rho$}}}

\newcommand{\grad}{{\mbox{\boldmath $\nabla$}}}

\newcommand{\lsim}{\mathrel{\hbox{\rlap{\lower.55ex \hbox{$ \sim$}} \kern-.3em \raise.4ex \hbox{$<$}}}}

\begin{document}

\title{Asymptotic results for backwards two-particle dispersion in a turbulent flow}

\author{Damien Benveniste${\,\!}^1$ and Theodore D. Drivas${\,\!}^{2}$}

\affiliation{ ${\,\!}^{1}$Department of Physics \& Astronomy \\
and ${\,\!}^{2}$Department of Applied Mathematics \& Statistics\\
The Johns Hopkins University, USA}

\date{\today}

\begin{abstract}
\noindent We derive an exact equation governing two-particle backwards mean-squared dispersion for both deterministic and stochastic tracer particles in turbulent flows. For the deterministic trajectories, we probe consequences of our formula for short time and arrive at approximate expressions for the mean squared dispersion which involve second order structure functions of the velocity and acceleration fields.  For the stochastic trajectories, we analytically compute an exact $t^3$ contribution to the squared separation of stochastic paths.  We argue that this contribution appears also for deterministic paths at long times and present direct numerical simulation (DNS) results for incompressible Navier-Stokes flows to support this claim.  We also numerically compute the probability distribution of particle separations for the deterministic paths and the stochastic paths and show their strong self-similar nature.
\end{abstract}
\maketitle



The dispersion of fluid particles backwards-in-time is critical for understanding the mixing properties of turbulent flows \cite{Sawford01}.  Mixing is a process by which scalar or vector quantities such as dye concentration, temperature or magnetic fields, are transported due to the action of an underlying flow \cite{Shraiman00,Bourgoin06}.  These fields are stretched and contorted, often violently, and if they have intrinsic diffusive properties (as most naturally occurring systems do), they are dissipated.  If the substance is non-diffusive, its value at a point is the initial value at the starting point of the {backwards deterministic trajectory} ending there.  The separation of these trajectories ending at nearby points is the principle subject of study for \emph{backward deterministic particle dispersion}.  For diffusive substances, initially separate bits of material can be brought together by the action of the turbulent flow and molecular effects.  Here an averaging process over {backwards stochastic trajectories} recovers the value at the field at any point.  The reversal of the process described above -- coalescence of forward perturbed trajectories -- is known as \emph{backwards stochastic particle dispersion}. Backwards particle dispersion is a key feature to understanding many physical problems ranging from hydrodynamic turbulence \cite{Eyink13}, passive scalar transport and dissipative anomalies \cite{Durbin80,Thomson03,Celani04}, cloud formation \cite{Bodenschatz10}, biological spread and enhanced  growth \cite{Garrett03,Garabato04,Visser07,Durham13}, air pollution \cite{Yassin12} stellar evolution \cite{Humphreys84,Ward02}, occurrence of solar flares \cite{Lazarian11} and galaxy formation \cite{Ryu08,Charbonnel07}.

It has been argued that in a turbulent flow, a pair of tracer particles separate proportionally to $t^{3}$ on average \cite{Richardson26}. This prediction is now known as the Richardson-Obukhov scaling law \cite{Obukhov41} and it remains a subject of controversy and inquiry. Experimental and numerical studies of two-particle dispersion are challenging as they require setting wide range separation between the dissipation scale and the integral scale of the flow \cite{Sawford01,Salazar09}. Backward dispersion is particularly difficult because it involves integrating the Navier-Stokes equation forward in time and tracking the tracers backward in time.  Most of the work on deterministic particle dispersion as therefore been in the forward-in-time setting. There the $t^3$ scaling has been numerically and experimentally thoroughly investigated though the physical mechanisms behind this growth are still not well understood. Furthermore, it has been observed that backward and forward dispersion are quite different \cite{Sawford05,Eyink12,Berg06}.   For example, particles are observed to spread faster backwards-in-time than forwards.  Since it is backward and not forward particle dispersion that matters for understanding mixing of passive fields, conclusive studies on scaling laws for backwards particle separations are of great importance.

In this paper we investigate the properties of backwards separation for both deterministic and stochastic tracers. We perform a systematic numerical study of deterministic tracers at Taylor-scale Reynolds number $Re = 433$. We investigate small and long time scaling laws, revealing a small time $t^4$ scaling and verifying the $t^3$ behavior at long time.  We also measure the observed Richardson constant for different final particle separations in addition to the corresponding probability distribution functions (PDFs) of the dispersion.  For the stochastic tracers, we analytically compute an exact $t^3$ term in the separation using Ito calculus and show numerically that this term is dominant at long times. This $t^3$ term seems to correspond to the asymptotic behavior of the deterministic tracers.

\vspace{-6mm}


\subsection{Deterministic Tracers}\label{dettracers}

\vspace{-4mm}
Consider a passive tracer with position $\bx(t)$ advected by the velocity field $\bu$ and whose second derivative is the acceleration field $\ba$.  For a Navier-Stokes velocity $\bu$, the $\ba$ is given by: $\ba\equiv \bF_{\rm ext}-\nabla p+\nu\triangle \bu$. The tracer is labelled at the final time $t_f$ and travels backwards for times $t\leq t_f$.  The dynamical relations are
\begin{align}
\dt\bx(t)= \bu(\bx(t)&,t), \ \ \bx(t_f)=\bx_f,\label{HE1}\\
\dt\bu(\bx(t),t)&= \ba(\bx(t),t).  \label{HE2}
\end{align}
Defining  $\tau(s)\equiv s-t$ and $\tau\equiv \tau(t_f)$, it follows from equations \eqref{HE1} and \eqref{HE2} that
\begin{align}\label{depth}
\bx(t) = \bx_f - \tau \bu(\bx_f,t_f)+ \int^{t_f}_t\tau(s) \ba(\bx(s),s)\rmd s.
\end{align}
The backward separation of particle pairs separated by $\br_f$ at the final time is $\br(\tau)\equiv \bx(t; \bx_f+\br_f)-\bx(t; \bx_f)$. The space-averaged squared separation satisfies:
\small
\begin{align}\nonumber
&\left\langle |\br(\tau)-\br_f|^2\right\rangle_{\bx_f}=  \tau^2 S_2^\bu(\br_f) +\left\langle\left| \int_t^{t_f}\tau(s) \delta \ba_L(\br_f; s)\ \rmd s\right|^2\right\rangle_{\bx_f}  \\ 
&\ \ \ \ \ \ \ \ \ \ \ \ -2\tau  \int_t^{t_f} \tau(s) \left\langle\delta \bu(\br_f; t_f)\cdot \delta \ba_L(\br_f;s)\right\rangle_{\bx_f}\rmd s,\label{squarepairs}
\end{align}
\normalsize
where $S_2^\bu(\br_f) =  \left\langle |\delta \bu(\br_f; t_f)|^2\right\rangle_{\bx_f}$ is the second order velocity structure function, $\delta \bu(\br_f;t_f)\equiv \bu(\bx_f+\br_f,t_f)-\bu(\bx_f,t_f)$ is the Eulerian increment of the velocity field at the final time,  $\delta \ba_L(\br;s)\equiv \ba(\bx(s;\bx_f+\br_f),s)- \ba(\bx(s;\bx_f),s)$ is the Lagrangian acceleration increment and $\langle\cdot \rangle_{\bx_f}$ denotes integration over the final particle positions ${\bx_f}$. This formula is exact -- no assumptions were needed to arrive at equation \eqref{squarepairs}.  The $\tau^2$ term appearing in equation \eqref{squarepairs} is the so-called Batchelor regime \cite{Batchelor50} where the particles undergo ballistic motion and is present for all times.  Note that in the forward case, the last term would have an opposite sign and $\{\bu_f , \br_f,t_f\}$ would be replaced by $\{\bu_0 , \br_0,t_0\}$.  If we consider small $\tau$, then a first order Taylor expansion of equation \eqref{squarepairs} yields:
\small
\begin{align}\label{short time}
\left\langle |\br(\tau)-\br_f|^2\right\rangle_{\bx_f} &\approx   S_2^\bu(\br_f)\tau^2 +2 \langle\varepsilon\rangle_\bx \tau^3 +\frac{1}{4}S_2^\ba(\br_f)\tau^4
\end{align}
\normalsize
where $S_2^\ba(\br_f)$ is the second order acceleration structure function at the final time.  We have also assumed that the final separation $\br_f$ is in the inertial range so as to use the Ott-Mann relation -- an (instantaneous) Lagrangian analogue of the $4/5$th law -- expressing $\left\langle\delta \bu(\br_f;t_f)\cdot \delta \ba(\br_f;t_f)\right\rangle_{\bx_f} \approx -2\langle \varepsilon \rangle_{\bx}$ where $\varepsilon$ is the viscous energy dissipation \cite{Falkovich01,OttMann00}. The $\tau^3$ term appearing in \eqref{short time} has been derived in \cite{Bitane12} for the forward case with an opposite sign and explored in DNS studies in \cite{Bitane13}. The equivalent term to our $\tau^4$ term has also been observed for the case of the velocity difference statistics \cite{Bitane13}. There it is described as the initial abrupt variation of the velocity difference and here we can further understand it as as a Bachelor-type-range (in a sense of first order expansion) for the velocity separation.

\vspace{-6mm}

\subsection{Stochastic Tracers}\label{stochtracers}
\vspace{-4mm}

\noindent   Consider now the following backward stochastic equation governing the flow of passive tracer particles:
\begin{align}\label{stochflow}
\rmd \tbx(t)=\bu( \tbx(t),t)\rmd t+\sqrt{2\kappa}\ \hat{\rmd}\bW_t, \ \ \  \tbx(t_f)=\bx_f.
\end{align}
Here $\kappa$ is the molecular diffusivity, $\bW_t$ is standard Brownian motion and $\hat{\rmd}$ is the backwards It\^{o}  differential \cite{Kunita90}.  Note that if the viscosity of the fluid is fixed and $\kappa\to 0$, we recover the deterministic equation \eqref{HE1}.  Along a path defined by equation \eqref{stochflow}, the backward It\^{o} lemma \cite{Kunita90,Constantin11} can be used to show that the Navier-Stokes velocity satisfies
\begin{align}\label{velstocha}
\rmd \bu(\tbx(t),t)=\ba^\kappa(\tbx(t),t)\rmd t+\sqrt{2\kappa}\nabla \bu|_{\tbx(t)}\cdot \hat{\rmd}\bW_t,
\end{align}
where $\ba^\kappa\equiv \bF_{\rm ext}-\nabla p+(\nu-\kappa) \triangle \bu$. This is a stochastic generalization of equation \eqref{HE2}. The quadratic variation term $-\kappa\Delta\bu$ appearing in the backward It\^{o} Lemma has the opposite sign than the forward case which would yield a similar expression but with $\ba^\kappa_{\rm forw} \equiv\bF_{\rm ext}-\grad p+(\nu+\kappa)\Delta\bu$. By integrating \eqref{velstocha}, we obtain
\begin{align}\label{posstocha}\nonumber
&\tbx(t) = \bx_f - \tau\bu(\bx_f,t_f)+ \int^{t_f}_t\tau(s)\ba^\kappa(\tbx(s),s)ds \\
&+ \sqrt{2\kappa}\int^{t_f}_t\tau(s)\grad \bu(\widetilde{\bx}(s),s)\cdot\rmd\bW_s - \sqrt{2\kappa}\bW_t.
\end{align}
This equation is the analogue of equation \eqref{depth} but for paths with intrinsic additive white noise.  Note that setting $\kappa=\nu$ leads to a dramatic simplification where the laplacian term in the acceleration vanishes: $\ba^\nu = \bF_{\rm ext}-\nabla p$.  Further, Eyink in \cite{Eyink12} has shown strong evidence that the backward dispersion is independent of $\kappa$ at long times. Motivated by its simplicity and the result of  \cite{Eyink12}, for the remainder of this work -- in both the theoretical and numerical results -- we make the choice of $\kappa=\nu$.  

Consider two trajectories $\tbx^1(t)$ and $\tbx^2(t)$ ending at the same point $\bx_f$ and satisfying equation \eqref{stochflow} with independent Brownian motions $\bW^1$ and $\bW^2$ respectively.  The natural object of study in this setting is the spaced averaged mean (in the sense of averaging over the Brownian motions) squared distance $\tbr(\tau)\equiv  \tbx^1(t_f-\tau)- \tbx^2(t_f-\tau)$ between the trajectories advected by independent Brownian motions.  In the deterministic case, in order to study the squared separation of two particles it was necessary to consider particles which (at the final time) were separated by a positive distance $|\br_f|$.  In the stochastic setting this is no longer necessary and starting particles at the same point effectively removes the dependence on the final separation and final velocity difference. 

 If the turbulence is homogeneous and isotropic or, more generally, the flow is ergodic, then the space averaged energy dissipation $\langle \varepsilon \rangle_\bx$ is constant in time.  Assuming this, at arbitrary time the space averaged mean dispersion can be proven to satisfy:
\begin{align}\label{disstocha}
\mathbb{E}_{1,2}\left\langle| \widetilde{\br}(\tau)+\sqrt{2\nu}\delta \bW_t|^2\right\rangle_{\bx_f}=\frac{4}{3}\langle\varepsilon\rangle_{\bx} \tau^3+ \mathcal{E}(\tau)
\end{align}
where the difference of the Brownian motions is denoted $\delta \bW_t \equiv \bW^1_t-\bW^2_t$ and where 
\begin{align}\label{Error}
\mathcal{E}(\tau)&=\mathbb{E}_{1,2}\langle\left|\delta\bA(\tau)\right|^2\rangle_{\bx_f}+2\mathbb{E}_{1,2}\left\langle\delta\bA(\tau)\delta\bU(\tau)\right\rangle_{\bx_f}.
\end{align}
The terms $\bA(\tau)$ and $\bU(\tau)$ are defined to be the integral terms involving $\ba^\nu$ and $\grad \bu$ respectively in equation \eqref{posstocha}.  The $\delta$ refers to the difference in a quantity evaluated on each path. The expectations are taken over realizations of the two independent Brownian motions.  The $\tau^3$ term is $\mathbb{E}_{1,2}\langle\left|\delta\bU(\tau)\right|^2\rangle_{\bx_f}$ after using It\^{o} isometry and the fact that $ \langle\varepsilon\rangle_\bx= \nu \langle |\nabla \bu|^2\rangle_\bx$.   If we instead decided to study forward stochastic tracers, then equation \eqref{disstocha} still holds but with minor differences in the error term including that integral $\bA(\tau)$ involve $\ba^\nu_{\rm forw}$ (defined earlier) instead of $\ba^\nu$.  Saffman in 1956 studied the effect of molecular diffusivity on forward particle dispersion using \eqref{stochflow} as a model \cite{Saffman60}.  There he computed short-time $t^3$ deviation of the stochastic dispersion to the deterministic for tracers in a homogenous isotropic turbulent flow.  The methods he employed are different from our own and, though his results have a similar form, they are not directly related to the long-time $4/3 \langle\varepsilon \rangle_\bx\tau^3$ behavior we calculate.

\vspace{-6mm}
\subsection{Numerical Results}
\vspace{-4mm}

In order to understand the behavior of the different terms in equation \eqref{disstocha}, we evaluate them using turbulence data from direct numerical simulations. We use the JHU Turbulence Database
Cluster \cite{Li08,JHU08}, which provides online DNS data over an entire large-eddy turnover time for isotropic and homogeneous
turbulence at Taylor-scale Reynolds number $Re_\lambda = 433$. The time history being stored online, it allows us to perform a brute force backward integration. Note that without this feature of the database, non trivial algorithms would need to be devised to overcome the storage problem \cite{Celani02}. The integration of particle trajectories is performed inside the database using the getPosition functionality \cite{Yu12} and a backward fourth-order Runge-Kutta integration scheme. 

\begin{figure}[!t] 
\begin{center}  
\includegraphics[width=\linewidth]{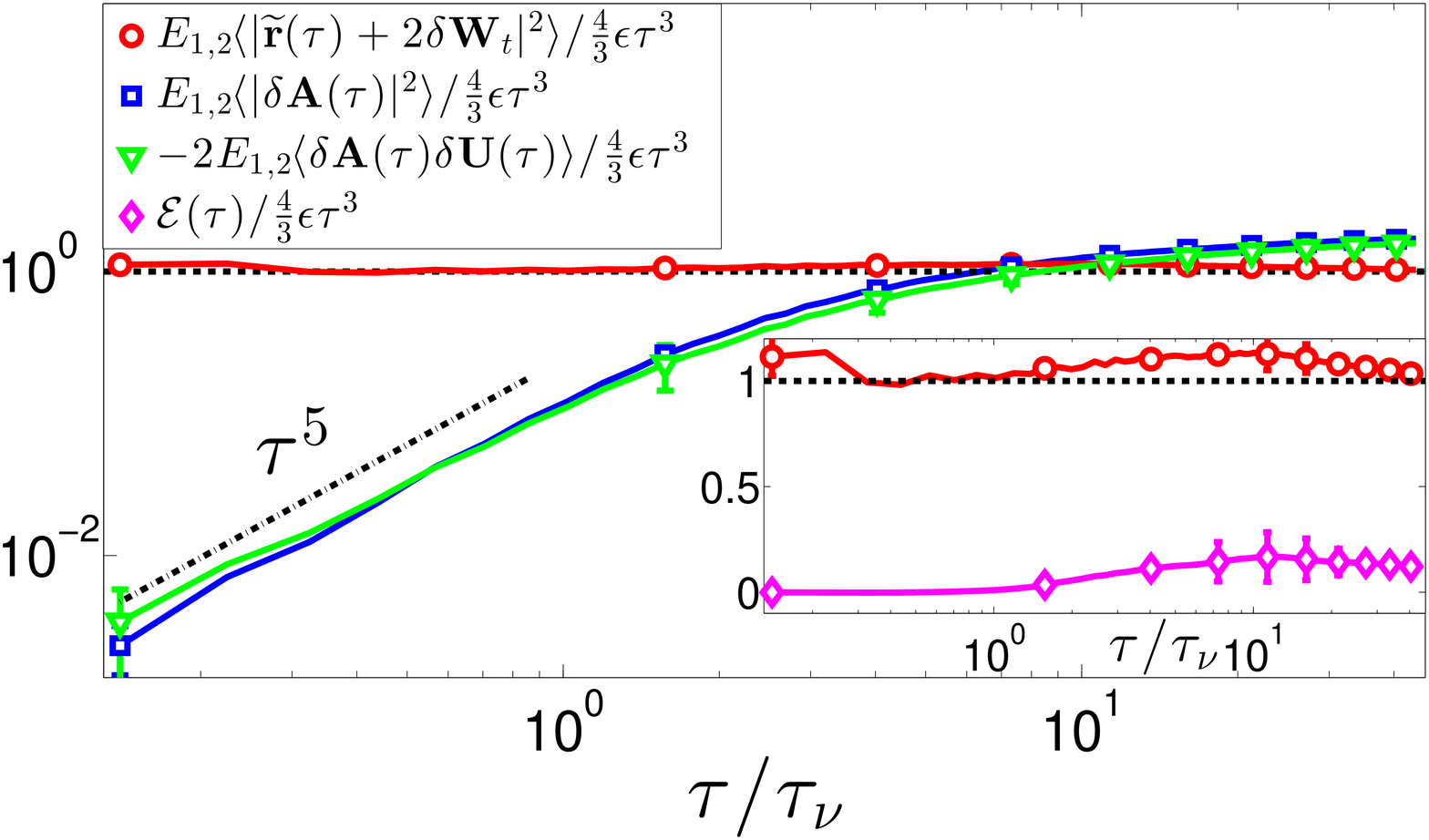}
\caption{Plots of the different terms of equations \eqref{disstocha} and \eqref{Error}. The dash-dotted line goes like $\tau^5$ and the dashed lines represent $\frac{4}{3}\langle\varepsilon\rangle_{\bx} \tau^3$. The inset is a closed-up view in semi logarithmic scale to compare the dispersion to $\mathcal{E}(\tau)$.} 
\label{Stochaplot} 
\end{center} 
\end{figure}

Figure \ref{Stochaplot} shows our results for the different terms in \eqref{disstocha} compensated by $4/3\langle\varepsilon\rangle_{\bx} \tau^3$. Error bars are calculated by the maximum difference between two subensembles of $N/2$ samples and $N=5\times10^8$. The terms $\mathbb{E}_{1,2}\langle\left|\delta\bA(\tau)\right|^2\rangle_{\bx_f}$ and $-2\mathbb{E}_{1,2}\left\langle\delta\bA(\tau)\delta\bU(\tau)\right\rangle_{\bx_f}$ behave in a very similar fashion. They grow as $\tau^5$ for short time and seem to reach an asymptotic $\tau^3$ for $\tau\gg\tau_\nu$, $\tau_\nu$ being the Kolmogorov time-scale. The difference  $\mathcal{E}(\tau)$ between these two terms however never exceeds $16\%$ of $4/3\langle\varepsilon\rangle_{\bx} \tau^3$ ($29\%$ including the error bars) at all points in time.  See inset of Figure \ref{Stochaplot}.  As a consequence of $\mathcal{E}(t)$ being small relative to $4/3\langle\varepsilon\rangle_{\bx} \tau^3$, we observe the dispersion is   
\begin{align}
\mathbb{E}_{1,2}\left\langle| \widetilde{\br}(\tau)+\sqrt{2\nu}\delta \bW_t|^2\right\rangle_{\bx_f}\approx\frac{4}{3}\langle\varepsilon\rangle_{\bx} \tau^3
\end{align}
for almost three decades.

We now take a moment to speculate on the behavior of particle trajectories for very high Reynolds number turbulence.  Note that, by the zeroth law of turbulence $\langle \varepsilon \rangle_\bx$ tends to a non-zero constant in the inviscid limit.  Therefore the $\frac{4}{3}\langle\varepsilon\rangle_\bx \tau^3$ term will remain for arbitrarily small viscosity.  Figure \ref{Stochaplot} shows that, in the inertial range, the terms $\mathcal{E}(t)$ is small compared to $\frac{4}{3}\langle\varepsilon\rangle_\bx \tau^3$.  In the inviscid limit, the inertial range expands to all scales and if $\mathcal{E}(t)$ is generically small in this range then there is good evidence for spontaneous separation of particle trajectories emanating from a single point in the zero viscosity limit.  This is the hallmark of the phenomenon of spontaneous stochasticity \cite{Bernard98,Gawedzki00,Chaves03,Kupiainen03,Vanden00,Vanden01,Chaves03,Eyink12}.   

\begin{figure}[t] 
\begin{center}  
\includegraphics[width=\linewidth]{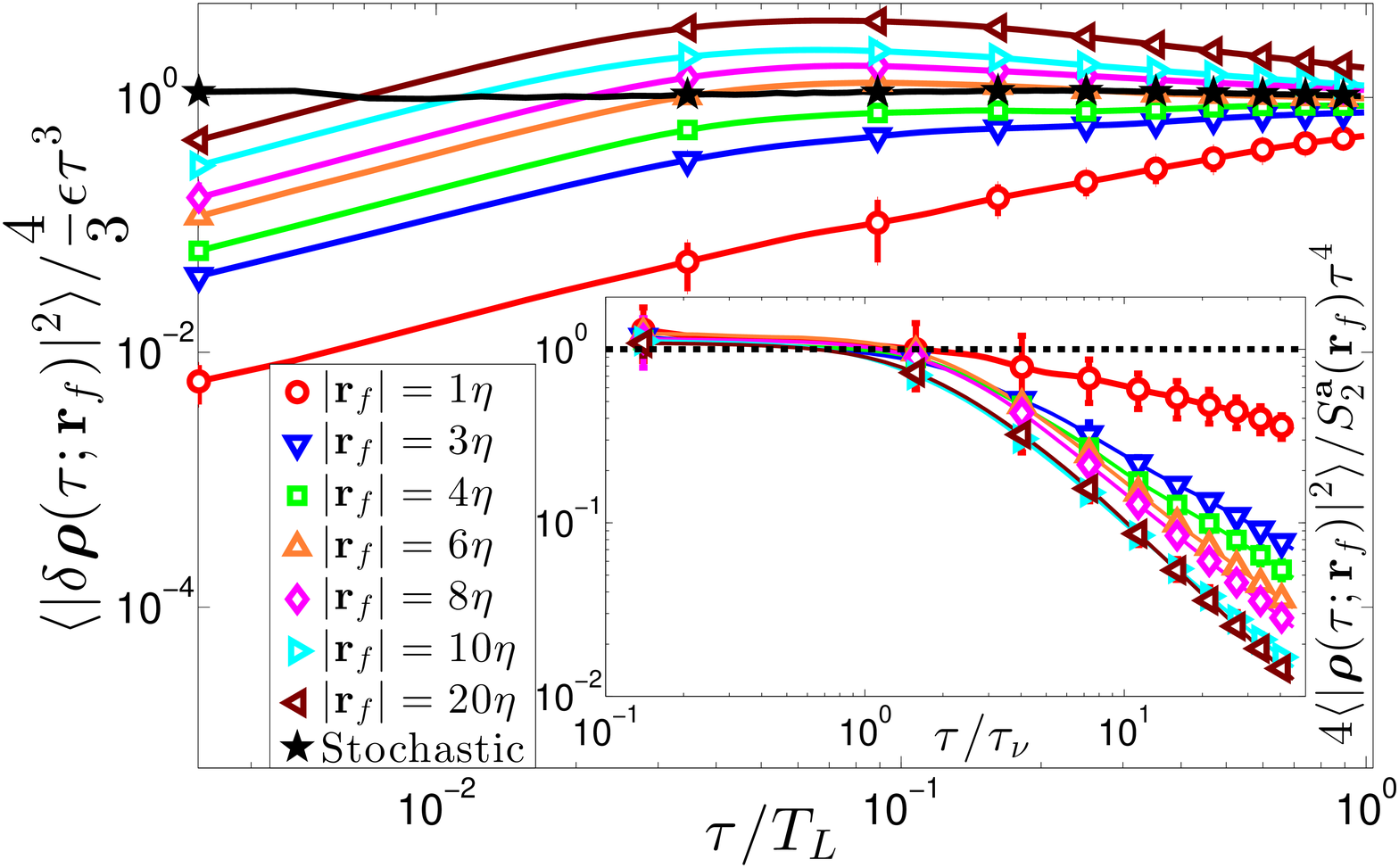}
\caption{Particle dispersion $\left\langle |\brho(\tau;\br_f)|^2\right\rangle_{\bx_f}$ for eight different initial separations in pure DNS data compared to the dispersion for the stochastic advection model. The curves are compensated by $\frac{4}{3}\langle\varepsilon\rangle_{\bx} \tau^3$. The inset shows the same pure DNS curves (without the stochastic case) rescaled by $\frac{\tau^4}{4}S_2^\ba(\br_f)$. The dashed line represents $\frac{\tau^4}{4}S_2^\ba(\br_f)$.} 
\label{rho} 
\end{center} 
\end{figure}

We now compare the stochastic particle dispersion of section \ref{stochtracers} to the deterministic dispersion of section \ref{dettracers} measured for finite final separations. We study the statistics of $\brho(\tau;\br_f)\equiv \br(\tau)-\br_f + \tau\delta \bu(\br_f;t_f)$.  By doing so, we effectively ``remove" the effect of the Bachelor regime and expose a longer range of $\tau^3$ scaling. It is equivalent to consider only the integral term over $\delta\ba(\br_f;s)$ in \eqref{squarepairs}. It was numerically verified that the quantity $\left\langle |\brho(\tau;\br_f)|^2\right\rangle_{\bx_f}$ has the asymptotic behavior 
\begin{equation}\label{limit}
\left\langle |\brho(\tau;\br_f)|^2\right\rangle_{\bx_f}\approx \begin{cases}
\left\langle | \br(\tau)|^2\right\rangle_{\bx_f}& \tau\gg\tau_\nu\\
\frac{\tau^4}{4}S_2^\ba(\br_f) &  \tau \lesssim  \tau_\nu
\end{cases}.
\end{equation}
 In Figure \ref{rho} we highlight the limiting cases given by \eqref{limit}.  The inset shows the $\frac{\tau^4}{4}S_2^\ba(\br_f)$ scaling law up to $\tau\approx\tau_\nu$ as predicted analytically.      In the main plot, results rescaled by  $\frac{4}{3}\langle\varepsilon\rangle_{\bx} \tau^3$ are shown for eight different final separations $r_f\in\left[\eta,20\eta\right]$ where $\eta$ is the Kolmogorov length scale and $T_L$ is the large scale turn-over time. The dispersion of the particles advected by the Brownian motion is in good agreement with the deterministic separations and all the curves tend to converge toward to the stochastic dispersion.  Defining $g_{\br_f}=\left\langle |\brho(T_L;\br_f)|^2\right\rangle_{\bx_f}/\langle\varepsilon\rangle_\bx T_L^3$, we record these measured ``Richardson constant" values in Table \ref{table:rich}.   The longest $\tau^3$ scaling (one decade) observed for deterministic dispersion is for $|\br_f|=6\eta$ and $g_{6\eta}=4/3$ within numerical error.  Note that in backwards setting, our measured Richardson constant is larger than forward results ($g\approx 0.5,0.52$) \cite{Salazar09,Bitane12,Bitane13} and is in good agreement with the measurements in  \cite{Sawford05,Berg06,Eyink12}.
 
\begin{table}[!b] 
\caption{Values of $g_{\br_f}$} 
\centering 
\begin{tabular}{|c| c |c |c|c| } 
\hline\hline 
$|\br_f|/\eta$& 4 & 6 & 8 &  \\ [0.5ex] 
\hline 
  ${g_{\br_f}}/{(4/3)}$& $0.86\pm 0.10$ & $0.99\pm 0.08$ & $1.17\pm 0.02$ &  \\ 
\hline\hline
$|\br_f|/\eta$ & 1 & 3 & 10 & 20 \\ [0.5ex] 
\hline 
  ${g_{\br_f}}/{(4/3)}$ & $0.50\pm 0.07$ & $0.76\pm 0.01$  & $1.24\pm 0.07$  & $1.71\pm 0.05$  \\ 
\hline 
\end{tabular} 
\label{table:rich} 
\end{table}

\begin{figure}[h] 
\begin{center}  
\includegraphics[width=\linewidth]{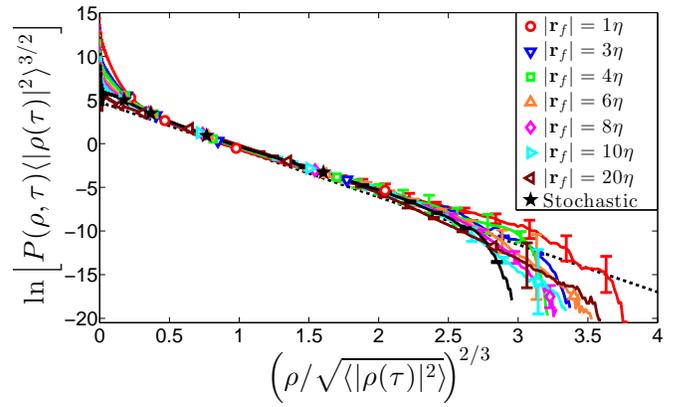}
\caption{Pair separation PDF for eight different initial separations in pure DNS data compared to the dispersion for the stochastic advection model with similarity scaling. Infinite Reynolds self-similar PDFs are shown for Richardson (straight dashed line).} 
\label{Proba} 
\end{center} 
\end{figure}

We have also numerically calculated the probability distribution functions (PDF) $P(\rho,\tau)$, $P(\wt{\rho},\tau)$ for the deterministic $\rho$ and stochastic $\wt{\rho}$ where $\rho\equiv \sqrt{\left\langle |\brho(\tau;\br_f)|^2\right\rangle_{\bx_f}}$and {\small $\widetilde{\rho}=\sqrt{\mathbb{E}_{1,2}\left\langle| \widetilde{\br}(\tau)+\sqrt{2\nu}\delta \bW_t|^2\right\rangle_{\bx_f}}$}.  Figure \ref{Proba} plots for $\tau=T_L$ the probability distributions with similarity scaling. The straight dashed line is the Richardson PDF:
\begin{equation} \label{richard}
P(\rho,t ) = \frac{B}{\langle\rho(\tau)^2\rangle} \exp\left[-A\left(\frac{\rho}{\langle\rho(\tau)^2\rangle^{1/2}}\right)^{2/3}\right].
\end{equation}
 All the curves are in good agreement with each others and Richardson for $0.4\lesssim(\rho/\sqrt{\langle\rho(\tau)^2\rangle})^{2/3}\lesssim2$. Note that $L$, the integral scale, we have $\left(L/\sqrt{\frac{4}{3}\langle\varepsilon\rangle_{\bx} T_L^3}\right)^{2/3}\simeq1.24$ in our case. Particle separations beyond that limit are outside the inertial range. Notice that the PDFs tends to spike above the Richardson PDF at small $\rho$ indicating for strong intermittency at small scales.

\begin{figure}[!t] 
\begin{center}  
\includegraphics[width=\linewidth]{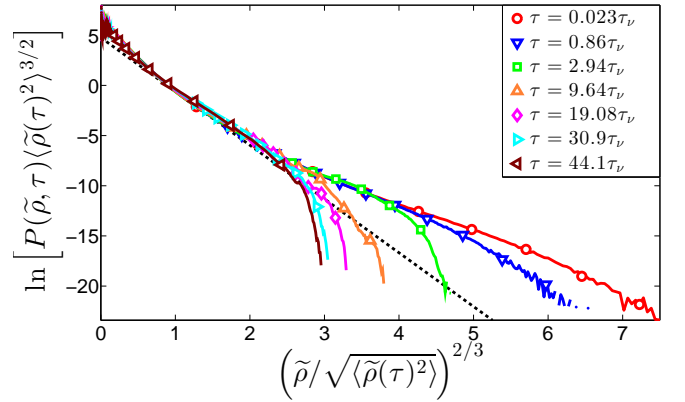}
\caption{Pair separation PDF for seven different times for the stochastic advection case with similarity scaling.The straight dashed line is the Infinite Reynolds self-similar PDF.} 
\label{Probatime} 
\end{center} 
\end{figure}

Figure \ref{Probatime} plots the PDF for the stochastic advection model at seven different times $\tau\in\left[0.023\tau_\nu,44.1\tau_\nu\right]$. The probability distribution exhibits a clear self-similarity behavior for $0\lesssim(\rho/\sqrt{\langle\rho(\tau)^2\rangle})^{2/3}\lesssim2$ at all times.   Note that even though we effectively removed the dependency of the dispersion on the final separation, the PDF is still not well described by Richardson \eqref{richard} at large scales (greater than 2 in similarity units).

We have investigated properties of mean squared dispersion for both deterministic and stochastic particle passive tracers in a turbulent flow.  We analytically predicted small time behavior for deterministic backwards dispersion and investigated the $\tau^4$-scaling law numerically.  We also looked at the convergence towards a long-time $\tau^3$ scaling law of the backwards dispersion for different final separations.  In addition, we developed a mathematical formalism for studying backwards stochastic particle trajectories with additive white noise.  Using this formalism we derived a formula for the mean dispersion and showed that the main contribution to this dispersion is a exact $4/3\langle\varepsilon \rangle_\bx \tau^3$ term which was calculated analytically (Figure \ref{Stochaplot}). We showed a striking agreement between the deterministic and stochastic cases for long times as all the dispersions seem to converge toward $4/3\langle\varepsilon \rangle_\bx \tau^3$ (Figure \ref{rho}).  Finally, we numerically compute the PDFs of particle separations in both the deterministic and stochastic settings.  In agreement with previous results for the forward case, the PDFs seem to be well-described by Richardson probability distribution for the same similarity units range.  As in the case of dispersion, the stochastic and deterministic case are in good agreement in this range.  Not only does the $4/3\langle\varepsilon \rangle_\bx \tau^3$ persist for all times, the PDF of the stochastic dispersion is self similar for all times up to 2 (in similarity units).  

There are a number of ways that the ideas in this paper could be further pursued.  First, we find it theoretically appealing that energy dissipation -- under mild assumptions -- can be related in a simple manner to the particle dispersion.  It would be interesting to probe the implications to the ``zeroth law of turbulence" from our framework. Next, in the present work, we make a particular choice of the noise strength which seems to be privileged for the case of backwards particle dispersion.  It would therefore be interesting to do a systematic study of the effect of changing the noise strength on the squared separation and probability distributions of the particles.  It would also be of great interest to use this formalism to -- for example -- predict features of multipoint statistical observables \cite{Frisch99}.  In principle one could study these quantities by using stochastic Lagrangian tracers and directly applying It\^{o} calculus methods coupled with numerics.

\vspace{-6mm}


\begin{acknowledgments}
\vspace{-3mm}

The authors would like to thank Gregory L. Eyink and Yi-Kang Shi for helpful discussions and suggestions.  This work was partially supported by NSF Grant No. CDI-II: CMMI 0941530 at Johns Hopkins University.

\end{acknowledgments}

  \bibliography{PREDamienTheo}

\end{document}